# SpectroGLY: A Low-Cost IoT-Based Ecosystem for the Detection of Glyphosate Residues in Waters

Javier Aira , Teresa Olivares , Francisco M. Delicado

*Abstract*—Glyphosate contamination in waters is becoming a major health problem that needs to be urgently addressed, as accidental spraying, drift or leakage of this highly water-soluble herbicide can impact aquatic ecosystems. Researchers are increasingly concerned about exposure to glyphosate and the risks its poses to human health, since it may cause substantial damage, even in small doses. The detection of glyphosate residues in waters is not a simple task, as it requires complex and expensive equipment and qualified personnel. New technological tools need to be designed and developed, based on proven, but also cost-efficient, agile and user-friendly, analytical techniques, which can be used in the field and in the lab, enabled by connectivity and multi-platform software applications. This paper presents the design, development and testing of an innovative low-cost VIS-NIR (Visible and Near-Infrared) spectrometer (called SpectroGLY), based on IoT (Internet of Things) technologies, which allows potential glyphosate contamination in waters to be detected. SpectroGLY combines the functional concept of a traditional lab spectrometer with the IoT technological concept, enabling the integration of several connectivity options for rural and urban settings and digital visualization and monitoring platforms (Mobile App and Dashboard Web). Thanks to its portability, it can be used in any context and provides results in 10 minutes. Additionally, it is unnecessary to transfer the sample to a laboratory (optimizing time, costs and the capacity for corrective actions by the authorities). In short, this paper proposes an innovative, low-cost, agile and highly promising solution to avoid potential intoxications that may occur due to ingestion of water contaminated by this herbicide.

## I. INTRODUCTION

Worldwide, glyphosate [N-(phosphonomethyl) glycine] is the systemic non-selective post-emergent herbicide most widely used to fight unwanted weeds [12], [13]. The use of glyphosate is regulated for the agricultural sector and for peri-urban and urban areas (albeit to a lesser extent in the latter). Regretfully, water, soil and food contamination with glyphosate is becoming a major problem that needs to be urgently addressed [14]–[26]. Glyphosate has been found in groundwater and surface water in several countries [27]–[31]. Raising awareness of the problem of glyphosate contamination is essential because, as it is highly soluble in water, it may impact aquatic ecosystems through accidental spraying, drift or leakage. Researchers [32]–[34] are increasingly concerned about the risks it poses for human health since, even in small amounts, agrochemicals may cause substantial damage and, as such, constitute a genuine public health problem. Agrochemicals may enter the human body through skin absorption, inhalation or intake. Symptoms in the ingestion of water contaminated with glyphosate (soluble concentrate), primarily gastrointestinal in nature, and including nausea, vomiting, diarrhea and belly pain, are associated with an intake of 5 up to 150 ml (they disappear within 24 hours). An intake of 20 up to 500 ml is associated with moderate gastrointestinal symptoms which last longer than 24 hours, such as gastrointestinal bleeding, esophagitis or gastritis verifiable via endoscopy, oral ulcers and hypotension. Serious intoxication (>500 ml) may cause respiratory failure, kidney failure, heart failure, coma and death [17]. The issue is highly complex and often difficult to quantify, due to the diversity of methods used to analyze exposure and also due to factor variability (including age, sex, nutritional status, personal habits and individual genetic variability), which affect, to a large extent, people's sensitivity to agrochemicals. One mechanism to avoid such exposure is the periodic analysis of waters to determine contamination levels and whether they are fit for use or not. This procedure requires taking samples, which are sent to a specialized laboratory. Complex equipment is used and operated by highly qualified personnel, while the results of the samples may take days to be published (processes that have significant costs).

In the context of laboratories, several analytical methods (traditional and complex) are used to determine the presence of glyphosate in waters, including EPA Method 547 [35], high performance liquid chromatography [36], mass spectrometry [37]–[40], fluorescence analysis [41], capillary electrophoresis analysis [42], spectrophotometric techniques [43]–[46], electrochemical techniques [47]–[49], and enzyme-linked immunosorbent assay [50], [51].

The global process of miniaturization, mass use and cost reduction in electronics and spectral sensors is in constant development.

This work was supported by the Spanish Ministry of Science and Innovation (FEDER, EU funds) under RTI2018-098156-B-C52 project.

Javier Aira is a PhD student at the International Doctoral School of the University of Castilla-La Mancha, 02071 Albacete, Spain (e-mail: jorgejavier.aira@alu.uclm.es).

Teresa Olivares is an Associate Professor with the Department of Computing Systems at the University of Castilla-La Mancha, 02071 Albacete, Spain (e-mail: teresa.olivares@uclm.es).

Francisco M. Delicado is an Associate Professor at the Department of Computer Engineering at the University of Castilla-La Mancha, 02071 Albacete, Spain (e-mail: francisco.delicado@uclm.es).

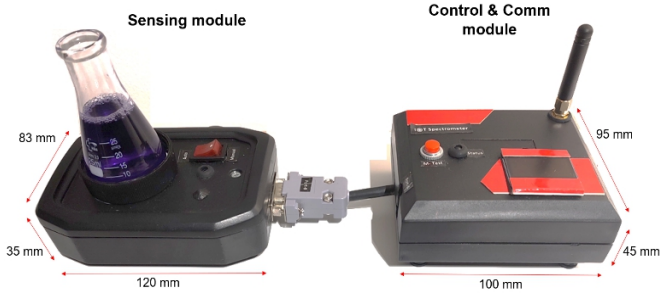

**Figure 1**. SpectroGLY with an analyzed water sample

The constant development of new electronic devices and spectral sensors of smaller size, cost, greater robustness and simplicity of use [1]-[11], is allowing the use of spectral techniques for the analysis of samples at very low cost and in portable systems. An example of this is the AS7265X multispectral sensor, which works in the spectral range from 410 to 940 nm and supports 18 wavelengths or channels [53].

This paper seeks to design and develop a VIS-NIR spectrometer, using an AS7565X sensor, based on IoT technology called SpectroGLY (Figure 1). The objective of SpectroGLY is to determine the presence of glyphosate in water, and to inform the user, through a digital traffic light, of the pollution level of this chemical substance. The digital traffic light establishes three levels: green, unpolluted water; yellow, warning level of glyphosate; and red, danger level (the chronic intoxications that may occur due to micro-ingestion sustained over time is outside the scope of this work). The user will be able to carry out the analysis of their samples in-situ, without the need to take them to the laboratory for analysis and in just 10 minutes. The main features of SpectroGLY include low cost, simplicity, portability, connectivity and integration with a Mobile App that interfaces with SpectroGLY via Bluetooth Low-Energy (BLE). It also has an IoT Platform, based on a micro-service architecture, which offers a Dashboard Web to view the results and statistics of tested samples. The purpose of this study is to propose an innovative solution for water safety and quality assurance.

The rest of the paper is structured as follows. Section 2 describes the system's global architecture, with special emphasis on the interactions between its several building blocks. Section 3 provides a detailed description of SpectroGLY. Section 4 contains a technical and functional description of the App associated with SpectroGLY. Section 5 describes the development of the IoT Platform's, where the information on the tests performed with SpectroGLY is concentrated, stored and displayed. Finally, Section 6 summarizes the conclusions and future works.

II. ARCHITECTURE IMPLEMENTED FOR SPECTROGLY

The primary purpose and challenge of the architecture was to take the current traditional lab measurement systems (benchtop spectrometers) to the next level, by offering a truly digital ecosystem that allows water samples (allegedly contaminated with glyphosate) to be tested using SpectroGLY, regardless of the measurement setting (rural or urban), and to ensure that results can be processed in a user-friendly and fast manner (Figure 2). In this respect, the architecture was conceived for its application in industry to be as accessible as possible, bridging the gap between the scientific and the industrial world, whilst considering all the digital services a user would require to ensure an outstanding and professional experience when using the system.

ThingsBoard was selected as the IoT Platform for this research study. This platform boasts a large community of developers and, technically speaking, has an interesting architecture based on micro-services and an embedded MQTT ("Message Queue Telemetry Transport") broker to support bi-directional communications from SpectroGLY, when it is based in urban areas and connected to a WiFi network. In this regard, for urban areas, bidirectional MQTT communication capabilities between SpectroGLY and the IoT Platform were also ensured, supporting standard telemetry and Remote Procedure Calls (RPCs), which allow commands to be sent from the Dashboard (IoT Platform) to the field equipment. The TLS ("Transport Layer Security") v1.2 protocol was implemented for greater security in all communications.

The IoT Platform supports integration with third party platforms, enabling integration with "The Things Networks" (TTN) middleware to achieve bi-directional communications when SpectroGLY is based in rural areas using a LoRaWAN. This network is part of the group of LPWANs [52], the popularity of which is growing in the IoT field due to their low-consumption and wide coverage range features. For this study, a LoRaWAN was implemented in the Australian band (AU915, 915-928 MHz), abiding by the radio spectrum standards issued by the Argentine federal telecommunications regulatory authority (ENACOM, in its Spanish acronym). Implementing a LoRaWAN requires several components. The Gateway is one of the most important and is in charge of spreading the network, establishing a bi-directional communication with such a network, and acting as a bridge to the TTN.

The Gateway used in this study was a Dragino LG02, which was integrated into the TTN through the HTTPS (Hyper Text Transfer Protocol Secure) protocol. The TTN was integrated into the IoT Platform through MQTT over TLS v1.2. In order to reinforce such integration, within the IoT Platform, there is a module called "Converters", in which two types of Converters are established—one for Uplink and the other one for Downlink. In the case of the Converter Uplink, the "Decoder" resource is used, which supports, via JavaScript, the parsing of data received from TTN (JSON or JavaScript Object Notation format). This allows us to display relevant information on the Dashboard, feed the Rule Engine and make such information persistent in databases, by allocating a specific type of variable, based on the type of data received. In the case of the Converter Downlink, the "Encoder" resource is used, which also supports JavaScript and allows the execution of the necessary instructions to convert JSON messages from TTN to commands, for instance, a Manual Test order, for SpectroGLY. These commands are submitted as LoRaWAN messages via the Gateway.

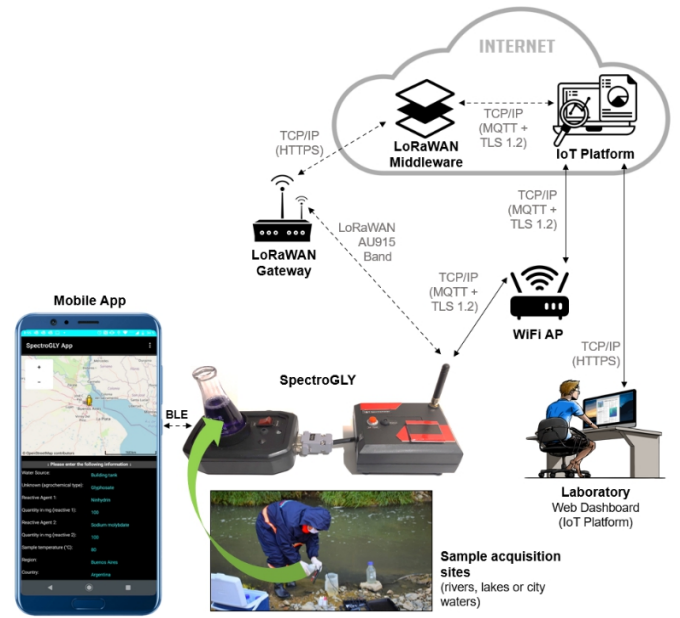

**Figure 2**. Digital ecosystem associated with SpectroGLY.

In extreme situations, where WiFi and LoRaWAN connectivity is unavailable, SpectroGLY could connect to a WiFi network propagated by a VSAT ("Very Small Aperture Terminal") to transmit data through the satellite link. The idea is to have multiple options for connecting to the cloud, with a device that is as transparent and standardized as possible to integrate it into any network without problems.

### III. SPECTROGLY

SpectroGLY is a substantially cheaper IoT device than traditional VIS-NIR spectrometers. For instance, the construction of the first SpectroGLY prototype required an investment of USD 250 (a ratio of approximately 10 to 1 vis-a-vis the most basic lab spectrometers in the market). In a serial production setting, this cost could decrease to approximately USD 200. Thanks to its reduced size, which makes it easily portable, SpectroGLY can be used both in the lab and in the field. In order to spotlight these characteristics, IoT technology was embedded to address connectivity concerns in those application settings, embracing digital tools, such as an App and Dashboard Web (IoT Platform), to view the results of the tests performed.

SpectroGLY supports two modes of use—manual and auto. The manual mode was only used to build the calibration curve, based on the reading of 12 water samples. To enable this mode, SpectroGLY has a mode selector to switch and enable the USB port (Port C) in order to connect it to a laptop, where it can be directly operated via a desktop software program called "AMS Spectral Sensor Dashboard" provided by the manufacturer of the multispectral sensor AS7265X [53]. This software is capable of reading data from the device's 18 channels. The software also provides controls for the reading update mode and LED controls (shutter), and has a section to observe calibrated raw data, including special functions to sort data by channel, wavelength and spectrum. After reading the samples required to build the calibration curve, the CSV ("Comma Separated Values") files were exported (one per each sample). These files were analyzed using the statistical software XLSTAT [54].

The auto mode will be the standard for use and, in principle, was devised to validate the model through test samples, incorporating into its firmware the linear regression equation of the model developed, as a result of the calibration curve construction. The auto mode requires the use of the App and Dashboard Web (IoT Platform) to view the results of the samples analyzed through a digital traffic light that will indicate whether or not the water samples contain glyphosate residues. SpectroGLY comprises two differentiated hardware modules: the sensing module, in charge of measuring the samples through the AS7265X multispectral sensor; and the communication & control module, which acts as a controller for the sensing module, integrates the IoT communication engines (WiFi, LoRaWAN and BLE), the power options (external through a conventional source and internal through a lithium battery for 24 hours of autonomy of continuous use) and provides for a set of sensors to ensure overall correct operation: temperature and humidity of the environment influence the accuracy of the AS7265X. In addition, a 3-axis accelerometer indicates whether the system is in a position to sample correctly. Figure 3 shows a detailed description of the implemented hardware.

At the user interface and for the auto mode, SpectroGLY has an OLED ("Organic Light-Emitting Diode") display that allows the user to be guided in the case of BLE connection problems with their smartphone (SpectroGLY App) or when connecting for the first time. In manual mode, the user is informed through the OLED display if there is a malfunction between the sensor and the laptop (USB). SpectroGLY also has a test button that allows test measurements to validate the correct operation of the device, its connectivity and the reception of data on the IoT platform. All SpectroGLY features are accompanied by a sound actuator (buzzer) that informs the user of any success or problems that may occur while performing their tests.

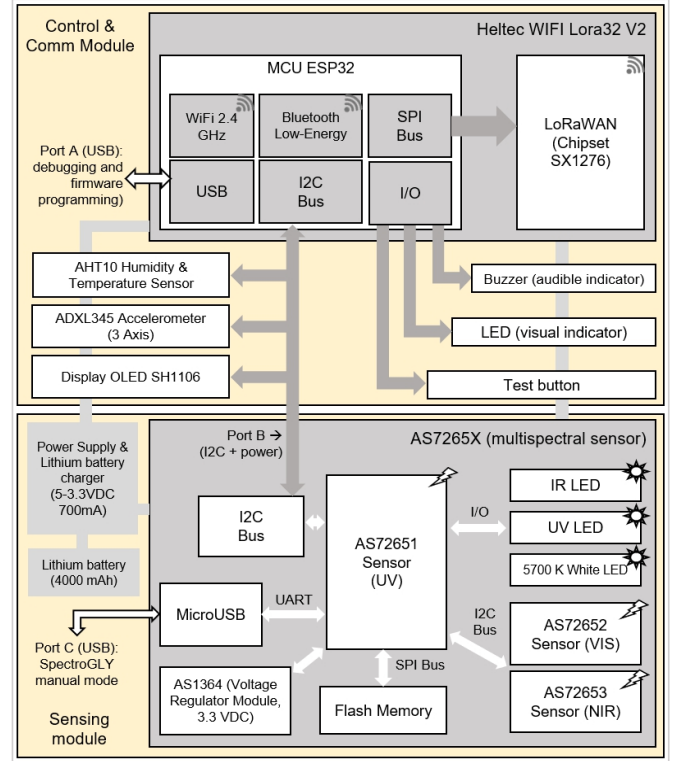

**Figure 3.** Implemented hardware.

*A. Sensing Module*

As mentioned in the preceding section, the sensing module comprises the multispectral sensor and supplementary electronics for integration with the communication & control module through the I2C ("Inter-Integrated Circuit") bus. The sensing module is in charge of sensing the water samples. SpectroGLY was based on the AS7265X multispectral sensor [53]. The AS7265X sensor has a spectral range from 410 to 940 nm, supports 18 wavelengths or channels (Figure 4), has a FWHM ("Full Width at Half Maximum") of 20 nm, comes ready calibrated, and has a programmable LED electronic shutter, with three (3) embedded LEDs as the external lighting source: 5700 Kelvin White LED, 405 nm Ultraviolet LED, and 875 nm Infrared LED.

AS7265X has interference filters embedded and directly deposited on CMOS silicon, temperature sensors and an MCU that supports bi-directional communications (I2C) with SpectroGLY's communication & control module. In order to validate the operation of the AS7265X sensor within SpectroGLY, for purposes of this research study, we decided not to perform calibration tests using conventional lab spectrometers. This decision was grounded in the large body of evidence from prior works conducted with the AS7265X multispectral sensor that empirically verify and confirm its correct operation [1]–[6]. Not only do these research studies provide evidence of the correct operation of the AS7265X sensor by comparing its performance and spectral response to lab spectrometers or similar devices, but they also succeed in comparing the spectral response of the sensor to high-resolution spectral libraries recognized in the academic world, as shown in the research conducted by AL-Qayssei et al. [4].

The analytical technique selected to determine the presence of glyphosate by SpectroGLY was that proposed by Besagarahally L. Bhaskara et al. [55]. Some changes were implemented to make water analysis processes more agile, lower the complexity and sample pre-processing times and, fundamentally, to minimize complexity in performing the analysis of samples. The aim of the latter was to expand the scope of potential users (the goal is for anyone with minimum technical training to be able to perform the tests). Naturally, the changes to the analytical method selected have no impact on its effectiveness, since the results were positive and highly promising (as will be discussed below). The analytical technique proposed by Besagarahally L. Bhaskara et al. is direct, simple, fast, accurate and fully consistent with the spectral portion supported by SpectroGLY (410 to 940 nm). It is based on a colorimetric chemical reaction of glyphosate in the presence of ninhydrin, a chromogenic reactive agent and the presence of a catalyzer, sodium molybdate dihydrate, which is easily detectable using diffuse reflectance spectroscopy. Such a reaction produces a violet product known as "Ruhemann's Purple" with maximum VIS absorption at 570 nm.

After defining the basic analytical technique used in determining glyphosate residues in water in SpectroGLY, a calibration curve was constructed (as it is done with lab spectrometers), based on 12 water samples fit for human consumption taken from the tap of a building apartment in the City of Buenos Aires, Argentina. Of the 12 samples, 4 revealed no presence of glyphosate (whites), while aliquots of soluble glyphosate ranging from 10 to 2000 mg/l were added to 8 samples. The concentrations of glyphosate mentioned above are reduced in a liter of drinking water. As a complement to the information on the health consequences of ingesting concentrated soluble glyphosate included in the introduction to this research work, it is worth noting that a dose of 10 mg/l does not present visible symptoms of intoxication in humans. However, coming into contact with a dose of this size leads to an increase in nuclear aberrations, which causes damage to DNA ("Deoxyribonucleic acid") [60]. Importantly, chronic and long-term consumption could cause health problems (an issue not addressed in this research work).

Figure 5 shows the results of the 12 water samples and the resulting colorimetric chemical reaction, according to the presence of aliquots of glyphosate, reactive agent, and catalyzer.

Table I shows the overall list of components used in this study and, in particular, the used quantities of ninhydrin and sodium molybdate dihydrate (100 mg in both cases). Initially, each of the 12 samples were pre-processed in 1000 ml beakers (borosilicate 3.3) under stable ambient conditions (temperature: 20° C and relative humidity: 55 %). After 10 minutes, the samples were quantitatively transferred to 25 ml Erlenmeyer beakers (borosilicate 3.3) for subsequent reading by SpectroGLY (manual mode).

After completing the 12 readings with SpectroGLY under manual mode and after obtaining the data associated with such readings (Table II), the CVS files related to each sample were exported from the "AMS Spectral Sensor Dashboard" software for subsequent processing by the XLSTAT statistical software in order to perform the linear regression equation of the model.

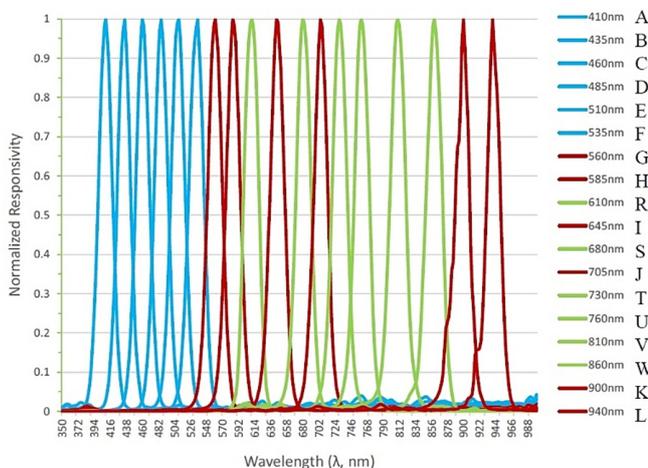

**Figure 4.** 18 wavelengths supported by the AS7265X sensor (410 to 940nm).

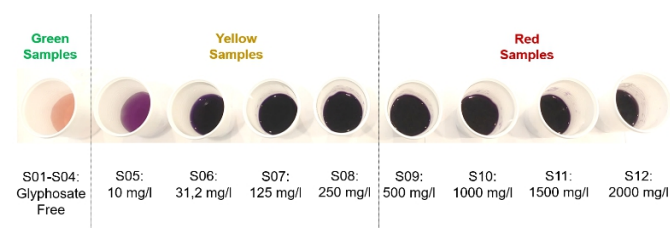

**Figure 5.** Colorimetric result obtained from the samples used for the construction of the calibration curve.

> REPLACE THIS LINE WITH YOUR MANUSCRIPT ID NUMBER (DOUBLE-CLICK HERE TO EDIT) <

5TABLE I
LIST OF COMPONENTS USED

| Item | Total Qty | Total Samples | Qty per sample (pre-process) | Composition / formulation | Qty per sample (analyzed) |
|---|---|---|---|---|---|
| City drinking water | 27 l | 27 samples (12: construction of the calibration curve & 15: test samples, validation model) | 1 l | $H_2O$ | 25 ml |
| Sodium molybdate dihydrate | 100 g | | 100 mg | $Na_2MoO_4 \cdot 2H_2O$ (reactive agent) | |
| Ninhydrine | 50 g | | | $C_9H_6O_4$ (reactive agent) | |
| Glyphosate Herbicide | 1 l | 18 samples (applicable to the study of samples with glyphosate) | Variable (mg/l) | Isopropylamine salt of N-phosphonomethyl glycine (48 g). Equiv. 36 g of acid glyphosate + Inert & adjuvants (100 ml) | |

After reading the samples using SpectroGLY and/or based on the data retrieved, a calibration curve was constructed, based on the glyphosate concentration within a range from 10 to 2000 mg/l. As mentioned, the technique proposed by Besagarahally L. Bhaskara et al. [55] produces a maximum VIS reaction at 570 nm. For this reason, during the calibration curve construction process, we had to choose a wavelength or channel to work with within SpectroGLY. To this end, the performance of wavelengths or channels from 560 nm to 585 nm was compared through a simple correlation matrix. After performing certain statistical analyses, we selected the 560 nm channel, without considering that this channel is 10 nm below that of the technique used by Besagarahally L. Bhaskara et al. (570 nm). As shown in Figure 6, through the spectral response obtained, we validated and confirmed that the selected analytical technique is feasible for use. Figure 7 shows the calibration curve carried out for the 560 nm wavelength or channel. Following the construction of the calibration curve that validated the performance of the selected wavelength, we were able to continue developing the linear regression equation of the model, based on the results obtained from the 12 samples summarized in Table II for this wavelength or channel.

TABLE II
TABLE WITH THE 12 READINGS OBTAINED (MANUAL MODE)

| Sample ID | Glyphosate Concentration (mg/l) | Traffic Light Result | Spectral response: Diffuse Reflectance (µW/cm2) | | | | | | | | | | | | | | |
|---|---|---|---|---|---|---|---|---|---|---|---|---|---|---|---|---|---|
| | | | 410 nm | 435 nm | 460 nm | 485 nm | 510 nm | 535 nm | 560 nm | 585 nm | 610 nm | 645 nm | 705 nm | 730 nm | 760 nm | 810 nm | 860 nm | 900 nm | 940 nm |

| Sample ID | Glyphosate Concentration (mg/l) | Traffic Light Result | 410 | 435 | 460 | 485 | 510 | 535 | 560 | 585 | 610 | 645 | 705 | 730 | 760 | 810 | 860 | 900 | 940 |
|---|---|---|---|---|---|---|---|---|---|---|---|---|---|---|---|---|---|---|---|
| S01 | 0 | Negative | 34 | 6 | 12 | 8 | 12 | 10 | 155 | 119 | 12 | 11 | 21 | 6 | 5 | 60 | 83 | 10 | 1 |
| S02 | 0 | Negative | 32 | 6 | 8 | 6 | 7 | 6 | 151 | 137 | 15 | 16 | 20 | 9 | 6 | 44 | 96 | 9 | 1 |
| S03 | 0 | Negative | 36 | 9 | 11 | 8 | 10 | 7 | 154 | 133 | 14 | 16 | 19 | 6 | 4 | 41 | 85 | 8 | 1 |
| S04 | 0 | Negative | 47 | 7 | 10 | 6 | 11 | 9 | 138 | 116 | 9 | 10 | 18 | 6 | 5 | 65 | 79 | 11 | 1 |
| S05 | 10 | Warning | 21 | 4 | 8 | 5 | 7 | 5 | 168 | 152 | 6 | 7 | 22 | 8 | 6 | 75 | 77 | 11 | 1 |
| S06 | 31,2 | Warning | 32 | 4 | 6 | 3 | 6 | 4 | 171 | 138 | 5 | 5 | 23 | 8 | 6 | 71 | 82 | 11 | 1 |
| S07 | 125 | Warning | 31 | 5 | 8 | 5 | 6 | 5 | 180 | 148 | 5 | 5 | 20 | 7 | 5 | 27 | 64 | 5 | 1 |
| S08 | 250 | Warning | 24 | 3 | 3 | 2 | 3 | 2 | 229 | 176 | 4 | 5 | 30 | 6 | 4 | 14 | 48 | 4 | 1 |
| S09 | 500 | Positive | 36 | 4 | 6 | 4 | 5 | 4 | 279 | 197 | 4 | 6 | 30 | 6 | 3 | 12 | 46 | 3 | 0 |
| S10 | 1000 | Positive | 32 | 4 | 5 | 3 | 4 | 3 | 297 | 256 | 7 | 7 | 34 | 6 | 3 | 10 | 42 | 4 | 0 |
| S11 | 1500 | Positive | 27 | 4 | 5 | 3 | 4 | 3 | 325 | 263 | 4 | 7 | 33 | 6 | 2 | 8 | 36 | 3 | 0 |
| S12 | 2000 | Positive | 32 | 4 | 4 | 3 | 3 | 3 | 375 | 306 | 4 | 8 | 37 | 5 | 2 | 8 | 34 | 4 | 0 |

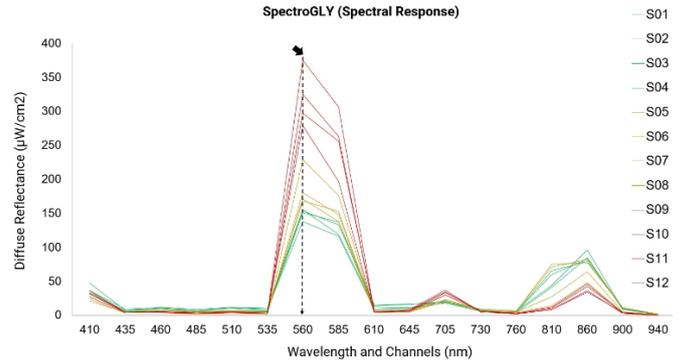

Figure 6. Spectral representation of the 12 samples analyzed (S1-S12).

In order to support the digital traffic light dimensioning within SpectroGLY for the established concentration range, we drew on a number of research studies [17]–[26] presenting several tests providing substantial empirical evidence of the clinical and statistical signs of mild, acute and chronic intoxication due to intake and inhalation of, or contact with, water contaminated with glyphosate. In said research, several studies were conducted on humans, animals and in vitro, reporting several indices that were taken as the reference to establish the traffic light range, according to the types of intoxication and their interaction with several sources of contaminated water. Some of the indices embedded in the traffic light correspond to an average lethal dose (DL50), average lethal concentration (CL50), NOAEL ("No-Observed-Adverse-Effect Level") and LOAEL ("Lowest-Observed-Adverse-Effect Level"). Considering the complexity involved in dimensioning a traffic light that would cover all problems associated with glyphosate contamination in waters, a global range from 10 to 2000 mg/l was established to evidence the correct operation of SpectroGLY and, within this range, establish the thresholds for the yellow light (Warning) and red light (Positive), based on the evidence, experiences and statistics observed in the reference research. The thresholds established for the yellow and red lights were also aligned with the extreme cases of acute intoxication due to direct intake, where symptoms are greater according to the amount ingested, and develop rapidly. Expanding the digital traffic light is a goal envisaged for future works, particularly, in the concentration range, to provide increased resolution in order for SpectroGLY to detect glyphosate at concentrations of below 10 mg/l.

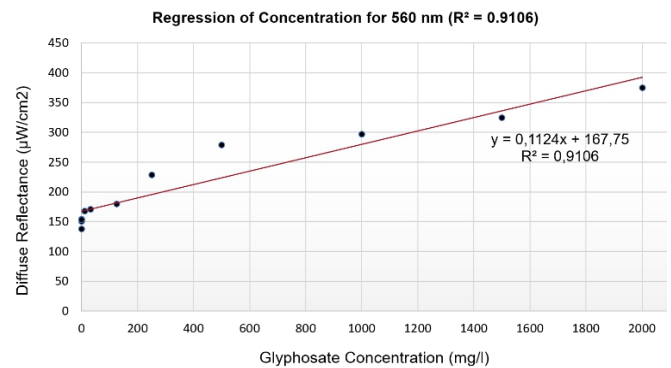

Figure 7. Calibration curve performed for the wavelength of 560nm.

To reinforce the evidence on the correct functioning of SpectroGLY, it was decided to add a laboratory spectrometer to this study. The AvaSpec-ULS3648 high-resolution spectrometer manufactured by Avantes [61] was used, with a spectral range of 200 to 1100 nm (VIS-NIR), a FWHM of 0.05 to 20 nm (configurable) and a lower precision error at 1%. The AvaSpec-ULS3648 is an entry-level laboratory spectrometer and costs around USD 2500 (ten times more expensive than SpectroGLY). To properly contrast the two instruments, the FWHM of the laboratory spectrometer was set to 20 nm, the spectral portion was limited from 410 to 940 nm, and the same 18 channels used in SpectroGLY were defined. As was done with SpectroGLY and using the same samples, the calibration curve for the laboratory spectrometer was constructed for the same wavelength used above (560 nm).

Table III shows the linear regression equations made for the SpectroGLY and laboratory spectrometer that allowed us to build the digital traffic light to determine the presence of glyphosate residues in water samples in both devices. The digital traffic light will show a green light when no glyphosate residues are found in the water sample analyzed (Negative). If the sample contains glyphosate residues (from 10 to 499 mg/l), a yellow light will be displayed (Warning), along with the recommendation of conducting more in-depth lab tests on the sample. Finally, if the sensed sample contains serious indications of the presence of glyphosate residues (from 500 to 2000 mg/l), a red light will be displayed (Positive).

To validate the correct functioning of SpectroGLY, 15 test samples were selected. Reactive agents (ninhydrin and sodium molybdate dihydrate) and their corresponding aliquots (100 mg each) were also added to the test samples. Regarding the reagents, in the future, no complications are foreseen in their handling and use, since the recommendation will be to have them available in individualized sachets with their corresponding ready-to-use aliquot. In this way, field tests will be carried out simply and without the need for the user to quantify and prepare the reagents (ready-to-use sachets will be available). Following the reading of the test samples, the auto mode was established in SpectroGLY, incorporating the linear regression equation of the model in its firmware, as a result of the construction of the calibration curve at 560 nm and the traffic light-specific parameters. At each reading of the test samples, we were able to empirically verify the correct operation of SpectroGLY, accurately allocating the traffic light

TABLE III
PARAMETERS OF THE DEVELOPED MODEL

Linear regression model equations:

SpectroGLY: Traffic Light value = [-1318,2455 + 8,0988 * X]

LAB Spectrometer: Traffic Light value = [-791,9610 + 6,1200 * X]

X = measured value at 560 nm

| Glyphosate Concentration (mg/l) | Traffic Light values [1] | | Traffic Light Result |
|---|---|---|---|
| | SpectroGLY | LAB Spectrometer | |
| Glyphosate Free | from -999,9999 (or less) to -62 | from -999,9999 (or less) to -57 | Negative |
| From 10 to 499 | from -61 to 537 | from -56 to 585 | Warning |
| From 500 to 2000 (or more) | from 538 a 999,9999 (or more) | from 586 a 999,9999 (or more) | Positive |

[1] Results of the model calibration and prediction studies

TABLE IV
RESULTS FROM THE TESTS SAMPLES

| Sample ID | Glyphosate Concentration (mg/l) | SpectroGLY | | LAB Spectrometer | | Traffic Light Result |
|---|---|---|---|---|---|---|
| | | Spectral response [1] | Traffic Light value | Spectral response [1] | Traffic Light value | |
| TS-01 | 0 | 90 | -589,3504 | 93 | -222,8020 | Negative |
| TS-02 | 0 | 100 | -508,3620 | 105 | -149,3620 | Negative |
| TS-03 | 0 | 117 | -370,6818 | 115 | -88,1620 | Negative |
| TS-04 | 0 | 135 | -224,9028 | 88 | -253,4020 | Negative |
| TS-05 | 0 | 145 | -143,9144 | 77 | -320,7220 | Negative |
| TS-06 | 15 | 169 | 50,4576 | 155 | 156,6380 | Warning |
| TS-07 | 50 | 175 | 99,0506 | 168 | 236,1980 | Warning |
| TS-08 | 150 | 184 | 171,9402 | 198 | 419,7980 | Warning |
| TS-09 | 200 | 218 | 447,3006 | 213 | 511,5980 | Warning |
| TS-10 | 220 | 224 | 495,8936 | 216 | 529,9580 | Warning |
| TS-11 | 600 | 285 | 989,9226 | 288 | 970,5980 | Positive |
| TS-12 | 800 | 290 | 1030,4167 | 293 | 1001,1980 | Positive |
| TS-13 | 1000 | 296 | 1079,0098 | 320 | 1166,4380 | Positive |
| TS-14 | 1600 | 337 | 1411,0620 | 357 | 1392,8780 | Positive |
| TS-15 | 2000 | 381 | 1767,4108 | 409 | 1711,1180 | Positive |

[1] Diffuse Reflectance (μW/cm2) for 560 nm

color to the samples analyzed (Table IV). As a complement to the results presented by SpectroGLY, Table IV also shows the results obtained with the laboratory spectrometer, where it is observed that the response of the digital traffic light for each of the 15 samples analyzed is the same as in SpectroGLY. Figure 8 shows the linear regression models and associated results with the test samples analyzed by both instruments. It can be seen that the SpectroGLY results are consistent and compatible with the laboratory spectrometer, and essentially no false positives or negatives occurred.

As a conclusion of the calibration and validation processes through the test samples, we empirically verified the correct performance of SpectroGLY in the detection of glyphosate residues in water samples. The selection of the AS7265X multispectral sensor was shown to be appropriate. Regarding the performance of SpectroGLY, the manufacturer of the multispectral sensor details that the accuracy error is 12 % (for the spectral channels expressed in μW/cm2) and the average aperture-limited field of view is ± 20.5° for the specified accuracy [53].

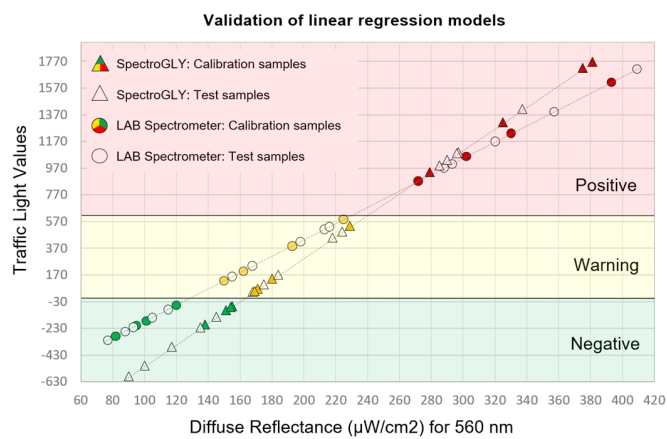

**Figure 8.** Graphic verification of the correct operation of the SpectroGLY.

*B. Communications & Control Module*

Within the SpectroGLY communications & control module (Figures 1 and 3), there is a Motherboard with a Heltec WiFi Lora32 V2 [56] development board, which contains a powerful 32-bit MCU based on ESP32 Tensilica 240 MHz LX6 dual-core [57]. This MCU supports a large computational capacity and many features. The communication & control module also offers several options in terms of IoT connectivity. It contains a SX1276-model LoRaWAN chipset (version 868 and 915 MHz), and a 2.4 GHz WiFi chipset (802.11 b/g/n 150 Mbps). This enables the transmission of events under the MQTT protocol and offers BLE v4.2 connectivity, thanks to which SpectroGLY is able to connect to the App from which the user controls and operates. All this is made possible by the large number of interfacing resources of the MCU.

*C. Firmware*

As mentioned in the preceding paragraph, SpectroGLY has a powerful MCU, thanks to which we were able to design, develop and express the operation logic, through a complex firmware using the C/C++ programming language. For the development of the firmware, the IDEs ("Integrated Development Environment") of Arduino and Visual Studio Code were used. To debug the developed firmware, the "ESP-Prog" hardware was used, which connects to the JTAG ("Joint Test Action Group") port of the ESP32 MCU to debug it in real time.

As mentioned, in terms of communications and for the urban context, the SpectroGLY firmware has the ability to connect and operate on WiFi networks (fixed or mobile). For this communications link, the MQTT protocol (over TLS 1.2) and a feature known as RPC were implemented in the SpectroGLY firmware, which allows information or orders to be received from the IoT Platform dashboard. The JSON data exchange format was used for the transmission and reception of information under the urban modality.

If SpectroGLY is deployed in a rural context, it is able to connect to LoRaWAN networks. It should be noted that the mechanisms required to send (Uplink) and receive (Downlink) messages were developed within the firmware. In the latter case, to receive information or orders from the IoT Platform dashboard (also using the TTN middleware). Due to the reduced size of the payload defined by the LoRaWAN standard for data messages, specifically for the frequency plan established and governed by ENACOM (AU915), the defined Payload ranges from 11 to 242 bytes. Therefore, we chose to use the "Cayenne Low Power Payload" (LPP) library, which provides a convenient and easy way to send data over LPWAN networks, such as LoRaWAN. Cayenne LPP [59] is a standardized and proven format that allows a substantial reduction in the number of bytes to be transmitted and allows SpectroGLY to send multiple data from its sensors and context parameters at the same time, in accordance with the rules and technical characteristics of SpectroGLY. In our specific case, using the Cayenne LPP library reduced the Payload to 148 bytes, including the global Cayenne LPP structure (Data ID + Data Type + Data Size).

To configure the technical parameters of both communication options and select the best option according to the operating context, SpectroGLY has the ability to receive the settings through the BLE link from the App, with said information persisting in its non-volatile Flash memory. From the same App, the user can set the specific parameters of the test to be performed (see next section), where SpectroGLY saves this information in its SRAM memory ("Static Random-Access Memory"). It then sends the information, adding the test results to the IoT platform. When the user asks SpectroGLY for a test of a water sample (via the App), the firmware makes a reading request to the AS7265X multispectral sensor (via the I2C Bus). After 15 seconds, the average time the sensor takes to send the response (metadata), the firmware processes this information and sends the result to the App to inform the user. A few seconds later, SpectroGLY sends all the test information to the IoT platform through the previously established connectivity.

IV. MOBILE APP

In conceiving, designing and developing SpectroGLY, the incorporation of a Mobile App was regarded as key for operating the device, and for obtaining real-time results from the water samples being analyzed. The tool selected for the App development was MIT App Inventor [58]. When the user runs the App for the first time, it must allow the use of two resources in particular: BLE and GPS. These resources are essential to communicate the App with SpectroGLY (BLE) and to establish the geopositioning (GPS) of the user. From a functional standpoint, the App developed for SpectroGLY allows the user to enter the required parameters to start running a water sample test. The parameters to be entered in the App include the source of the water sample (river, lake, city water, etc.), type of agrochemical suspected to be present in the sample (in this case, glyphosate), types of reactive agents and their aliquots, and context data (country and region, city or province). After entering the basic parameters required by the system to analyze the sample, the user will execute the sample read request to SpectroGLY. Once the user has requested the reading, and after 15 seconds, SpectroGLY will then send the test result to the App. After notifying the sample test results, SpectroGLY will also provide context information to contribute to the sample traceability (timestamp, test identifier number, SpectroGLY serial number, and IoT communication link). As supplementary functionalities, the App also offers the user configuration options in SpectroGLY, specifically, the ability to change communication parameters in the IoT Platform. For example, the ability to change and request WiFi network parameters in SpectroGLY (establishing a new SSID and its associated password) and, if the device is in a rural setting and there is LoRaWAN coverage, the ability to easily establish such connectivity. Finally, the App also informs the user of whether the data has been successfully transmitted to the IoT platform.

V. IOT PLATFORM

As mentioned, for this study, we selected ThingsBoard as the IoT Platform. ThingsBoard supports the collection, processing, visualization and management of generic and customized IoT devices. Additionally, it combines scalability and performance to ensure data persistence. ThingsBoard allows IoT entities to be provided, overseen and monitored in a secure manner, through APIs enriched from the server end. ThingsBoard allows relationships to be established across devices, assets, customers or any other entity. The platform is able to transform and standardize data from the devices, may establish and generate incoming telemetry event alarms, attribute updates, device downtime, and user's actions. For this study, a project was created within ThingsBoard. SpectroGLY was integrated through the MQTT protocol for the urban mode, where the device has access to WiFi networks and supports bidirectional communications: SpectroGLY is able to send information to the IoT Platform, as well as to receive messages from it through RPC. Information is thus displayed on the Dashboard, the Rule Engine is fed, and such information persists in the databases. For the rural setting, the LoRaWAN TTN middleware was integrated through the MQTT protocol in a bidirectional manner, supporting the Uplink and Downlink methods. The Dashboard displays the SpectroGLY geo-positioning, based on GPS coordinates from the user's smartphone (App), the spectral signature obtained from the tested samples and sample results. It provides the ability to perform manual tests and view the most significant parameters of each test: test number, sample ID, sample source, agrochemical analyzed, type/quantity of reactive agents used, and other parameters that help better understand the context in which the test was run. As a complement, the platform also validates and ensures that the measurements in the field are carried out at ambient temperature and humidity. For temperature, allowable measurements range from 20° to 24° C and for humidity, accepted measurements range from 40 % to 70 %. The position of the equipment at the time of measurement is also analyzed, thanks to the position sensor (accelerometer) integrated in SpectroGLY. All the information sent from SpectroGLY persists on databases within the IoT Platform and is available for the user to execute searches by date and time, ensuring the respective traceability.

## VI. Conclusions

In conclusion, we designed, developed, calibrated and verified the correct operation of SpectroGLY, a low-cost hand-held device based on VIS-NIR spectrometry, customized to detect glyphosate residues in waters in rural and urban settings. Thanks to its compact design, it may be used in the lab and in the field. Such flexibility was made possible by using IoT technology which, in addition to the hardware we designed and developed (with embedded IoT connectivity), also incorporates an App and Dashboard Web (IoT Platform) to offer the user a true and complete digital ecosystem to see the results and statistics of measurements made from a smartphone or laptop.

Thanks to SpectroGLY, the user of the system will be able to obtain the results of their samples in 10 minutes, avoiding said samples having to be transferred to the laboratory for their evaluation, optimizing times and costs. Early warnings can also be issued to the pertinent authorities so that these may implement corrective actions to prevent people from having access to water contaminated with glyphosate, thus avoiding intoxications.

As regards future work, the technology developed could be adapted and evolved into a 100 % stand-alone IoT sensor that will allow users to take water samples at a given setting through a small water pump, pre-process sample (inject them with aliquots of activating reactive agents), run the tests, and finally send the results to the IoT Platform by means of the connectivity available in the measurement setting.

## Acknowledgment

Grant PID2021-123627OB-C52 funded by MCIN/AEI/10.13039/501100011033 and by "ERDF A way to make Europe". Our recognition and gratitude to the International Doctoral School of the University of Castilla-La Mancha for their support in conducting this research work.

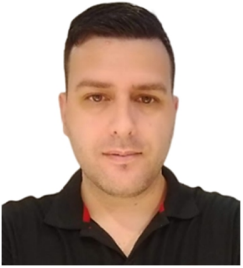
**Javier Aira** was born in the Autonomous City of Buenos Aires, Argentina, in 1982. He obtained a Master's Degree in Strategic Management in Information Technology from the European University of the Atlantic (Santander, Spain). He has a degree in Management of Automation and Robotics Systems from the National University of Lomas de Zamora (Buenos Aires, Argentina). He is currently a PhD student at the University of Castilla-La Mancha (Albacete, Spain) in Advanced Computer Technologies in the IoT discipline in Smart Cities. He has more than 17 years of experience in the industry, specifically in the area of R&D and in projects related to the Internet of Things.

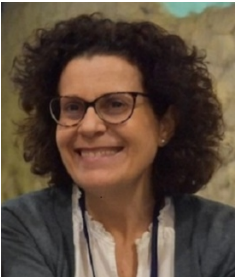
**Teresa Olivares** is an Associate Professor with the Department of Computing Systems at the University of Castilla-La Mancha. She received her PhD degree in Computer Science in 2003 from the same university. She is a member of the research group High-Performance Networks and Architectures at the Albacete Research Institute of Informatics. Her main scientific research interests include Internet of Things standards, communications and protocols, heterogeneous low power wireless sensor networks and standards, smart environments, Industry 4.0 and Reverse Logistics. She has participated in more than 40 research projects and has co-authored more than 50 research papers in journals, conferences and book chapters.

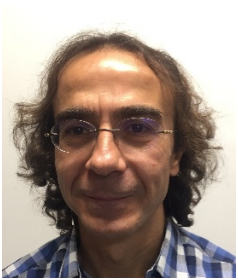
**Francisco M. Delicado** received his PhD degree in Computer Engineering in 2005 from the University of Castilla-La Mancha. He has been an Associate Professor at the Department of Computer Engineering at this university since 2007 and member of the research group High-Performance Networks and Architectures at the Albacete Research Institute of Informatics. His research interests include SDN (Software Defined Networking), WSN (Wireless Sensor Networks), heterogeneous low power WSN and cloud networking.